\documentclass[aps,prl,twocolumn,showpacs,preprintnumbers]{revtex4}
\usepackage{amsmath,amssymb}
\usepackage{bm}
\usepackage{graphicx}
\usepackage{comment}
\newcommand\x{\mathbf{x}}
\newcommand\y{\mathbf{y}}
\newcommand\z{\mathbf{z}}
\newcommand\p{\mathbf{p}}
\newcommand\q{\mathbf{q}}
\renewcommand{\d}{\partial}
\begin{document}
\preprint{INT-PUB-11-010}
\title{Berry Curvature, Triangle Anomalies, and the Chiral Magnetic 
Effect in Fermi Liquids}
\author{Dam Thanh Son and Naoki Yamamoto}
\affiliation{Institute for Nuclear Theory, University of Washington,
 Seattle, Washington 98195-1550, USA}
\begin{abstract}
In a three-dimensional Fermi liquid, quasiparticles near the Fermi
surface may possess a Berry curvature.  We show that if the Berry
curvature has a nonvanishing flux through the Fermi surface, the
particle number associated with this Fermi surface has a triangle
anomaly in external electromagnetic fields.  We show how Landau's
Fermi liquid theory should be modified to take into account the Berry
curvature.  We show that the ``chiral magnetic effect'' also
emerges from the Berry curvature flux.

\end{abstract}

\maketitle

\emph{Introduction.}---%
Recently there has been a lot of interest
in the effect of the Berry phase and Berry curvature on the physics of the
electron Fermi liquid.  The standard theory of Fermi liquids,
developed by Landau~\cite{Landau}, assumes that the low-energy degrees
of freedom in a Fermi liquid are the fermion quasiparticles, whose
distribution function in phase space satisfies a kinetic equation.  In
many cases the semiclassical motion of a wave packet of electrons in a
crystal should include an extra term due to the Berry phase,
expressible in terms of the electronic Bloch wave
functions~\cite{SundaramNiu}.  Such a term should alter the standard
kinetics of Fermi liquids (see also below); including this term leads
to an interpretation of the anomalous Hall conductivity in terms of
Fermi surface properties~\cite{Haldane}.

In this Letter we show the connection between the Berry curvature on
the Fermi surface and triangle anomalies.  Let us first notice that
the total flux of Berry curvature through a given Fermi surface does
not need to vanish, but can be a multiple of the flux quantum.  One
possible case is doped Weyl semimetals~\cite{Vishwanath,BurkovBalents,Xu-chern},
but the discussion does not depend on the origin of the Berry
curvature flux.  We will show that if there are $k$ quanta of Berry
curvature flux through a given Fermi surface, then the number of
particles associated with this Fermi surface (which is proportional to
its volume) is not conserved in the presence of the external
electromagnetic field,
\begin{equation}\label{dn-anom}
  \frac{\d n}{\d t} + \bm{\nabla}\cdot{\bf j} = \frac k{4\pi^2}
  {\bf E}\cdot {\bf B}.
\end{equation}
Charge conservation is ensured by the vanishing of the sum of $k$'s of
all Fermi surfaces.

Equation~(\ref{dn-anom}) is exactly the equation of triangle anomalies
in relativistic quantum field theory~\cite{Adler,BellJackiw}.  It is
therefore expected to hold for a Fermi gas of relativistic fermions at
finite density.  Indeed, the Berry curvature of a relativistic fermion
has the form of the field of a magnetic monopole in momentum space, and
$k=\pm1$ for right- (left-)handed fermions.  The
statement~(\ref{dn-anom}) goes further by tying anomalies with Fermi
surface properties only.  In this way, we demonstrate that axial
anomalies are properties of Fermi liquids with Berry curvature flux,
even when the original particles interact strongly.  
The only assumption is that low-energy degrees of freedom are
fermions that are described by Landau's Fermi liquid theory; 
the extension to, e.g., the superfluid A phase of $^3$He \cite{Volovik}
is deferred to a future question, where Nambu-Goldstone bosons associated 
with the spontaneous breaking of U$(1)$ particle number must also be 
taken into account. (In such a Weyl superfluid, the relation between the
topology of a Fermi surface and anomalies was studied theoretically in a 
different way and verified experimentally;
see Ref.~\cite{Volovik} and references therein.)

As is evident from our arguments, triangle anomalies in a Fermi liquid have
a ``kinematic'' origin, independent of the details of the Hamiltonian.
Namely, we will show that Berry curvature modifies the commutation
relation of the particle number density operator, and that this
commutator is related to the anomalous Hall effect and the triangle
anomalies for the fermion numbers near one Fermi surface.

\emph{Hamiltonian formulation of Landau's Fermi liquid theory.}---%
The fundamental equation of Landau's Fermi liquid theory is a kinetic
equation governing the time evolution of the occupation number of
quasiparticles $n_\p(\x)$,
\begin{equation}\label{kin-eq-FL}
  \frac{\d n_\p(\x)}{\d t} 
  + \frac{\d\epsilon_\p}{\d\p}\cdot
  \frac{\d n_\p}{\d\x} 
  - \frac{\d\epsilon_\p}{\d\x}\cdot \frac{\d n_\p}{\d\p}=0,
\end{equation}
where $\epsilon_\p=\epsilon^0_\p+ \delta\epsilon_\p$, $\epsilon^0_\p$
is the energy of a single quasiparticle excitation with energy $\p$,
and $\delta\epsilon_\p$ is the modification of its energy due to
interactions with other quasiparticles,
\begin{equation}
  \delta\epsilon_\p = \int\!\frac{{\rm d}\q}{(2\pi)^3}\,
  f(\p,\q)\delta n_\q(\y),
\end{equation}
$\delta n_\q(\y)=n_\q(\y)-n_\q^0$ is the deviation from the ground
state distribution function and $f(\p,\q)$ are Landau's
parameter. Above we have neglected the collision term.

For the purpose of generalizing Landau's Fermi liquid theory to
systems with Berry curvature, we reformulate the kinetic equation as
the evolution equation of a Hamiltonian system.  In this formulation,
the kinetic equation has the form
\begin{equation}\label{H-evol}
  \d_t n_\p(\x) = {\rm i} [H,\, n_\p(\x)],
\end{equation}
where the Hamiltonian $H$ is the conserved energy,
\begin{equation}\label{H-FL}
  H = \int\!\frac{{\rm d}\p\,{\rm d}\x}{(2\pi)^3}\, \epsilon^0_\p\delta n_\p +
  \frac12 \int\! \frac{{\rm d}\p\,{\rm d}\q\,{\rm d}\x}{(2\pi)^6}\, 
  f(\p,\,\q) \delta n_\p \delta n_\q,
\end{equation}
and the commutator is postulated as
\begin{multline}\label{comm-FL}
  [ n_\p(\x),\,  n_\q(\y)] = -{\rm i}
  (2\pi)^3 \frac{\d}{\d\p}\delta(\p-\q)\cdot\frac{\d}{\d\x}\delta(\x-\y)\\
  \times [n_\p(\y)-n_\q(\x)].
\end{multline}
It is straightforward to verify that Eqs.~(\ref{H-evol}), (\ref{H-FL}), and
(\ref{comm-FL}) imply Eq.~(\ref{kin-eq-FL}).

The commutation relation~(\ref{comm-FL}) is remarkable in the
following respect: assume we have two operators, $\hat A$ and $\hat
B$, linear in occupation numbers,
\begin{equation}
  \hat A =\! \int\!\frac{{\rm d}\p\,{\rm d}\x}{(2\pi)^3}\, 
    A(\p,\x) n_\p(\x),~~
  \hat B =\! \int\!\frac{{\rm d}\p\,{\rm d}\x}{(2\pi)^3}\, B(\p,\x) n_\p(\x),
\end{equation}
then its commutator will be
\begin{equation}
  [\hat A, \, \hat B] = -{\rm i}\!\int\!\frac{{\rm d}\p\,{\rm d}\x}{(2\pi)^3}\,
   \{A,\, B\} n_\p(\x),
\end{equation}
where $\{A,\, B\}$ is the classical Poisson bracket
\begin{equation}
  \{A,\, B\}(\p,\x) = \frac{\d A}{\d\p} \cdot \frac{\d B}{\d\x} -
  \frac{\d A}{\d\x} \cdot \frac{\d B}{\d\p}\,.
\end{equation}
The presence of the Berry curvature, as we shall see, changes the
classical Poisson bracket and leads to a modification of Landau's
Fermi liquid theory.

\emph{Berry curvature and Poisson brackets.}---%
Before tackling the many-body physics of Fermi liquids, let us
consider a single quasiparticle in a theory with Berry curvature of
the Fermi surface.  Such a quasiparticle is described
by the action~\cite{Xiao:2005,Duval:2005}
\begin{equation}\label{S-Berry}
   S = \int\!{\rm d}t\, 
  \left[p^i \dot x^i + A_i(x)\dot x^i - {\cal A}_i(p)\dot p^i - H(p,x)\right],
\end{equation}
where $H(p,x)$ is the Hamiltonian whose form is not important for us
right now, $A_i$ is the electromagnetic vector potential, and ${\cal
A}_i(p)$ is a fictitious vector potential in momentum space.  
Combining $p$ and $x$ into one set of variables $\xi^a$,
$a=1,\ldots,6$, the action can be written as
\begin{equation}
  S =\!\int\!{\rm d}t\, [ -\omega_a(\xi)\dot\xi^a - H(\xi) ].
\end{equation}
The equations of motion that follow from this action are
\begin{equation}
  \omega_{ab} \dot\xi^b = - \d_a H \,,
\end{equation}
where $\omega_{ab} = \d_a\omega_b-\d_b\omega_a$ and
$\d_a\equiv\d/\d\xi^a$. We can reinterpret this equation as
\begin{equation}
  \dot\xi^a = \{H,\, \xi^a\} = - \{\xi^a,\, \xi^b\} \d_b H \,,
\end{equation}
where the Poisson bracket is defined as
\begin{equation}
  \{\xi^a,\, \xi^b\} = (\omega^{-1})^{ab} \equiv \omega^{ab},
\end{equation}
where $\omega^{-1}$ is the matrix inverse of $\omega_{ab}$.  For the
action~(\ref{S-Berry}), the Poisson brackets are~\cite{Duval:2005}
\begin{subequations}\label{PB}
\begin{align}
  \{ p_i,\, p_j\} &= - \frac{\epsilon_{ijk} B_k}{1+{\bf B}\cdot\bm{\Omega}}\,, \\
  \{ x_i,\, x_j\} &= \frac{\epsilon_{ijk}\Omega_k}{1+{\bf B}\cdot\bm{\Omega}}
   \,,\\ 
  \{ p_i,\, x_j\} &= \frac{\delta_{ij} + \Omega_i B_j}
  {1+{\bf B}\cdot\bm{\Omega}}\,,
\end{align}
\end{subequations}
where $B_i=\epsilon_{ijk}\d A_k/\d x_j$, $\Omega_i=\epsilon_{ijk}\d
{\cal A}_k/\d p_j$.

The invariant phase space is (here $\omega\equiv\det\omega_{ab}$) \cite{Xiao:2005}
\begin{equation}\label{phase-space}
 {\rm d}\Gamma = \sqrt{\omega}\, {\rm d}\xi
  =(1+ {\bf B}\cdot\bm{\Omega}) 
  \frac{{\rm d}\p\,{\rm d}\x}{(2\pi)^3}\,.
\end{equation}

It is now clear how to incorporate Berry curvature into Landau's Fermi
liquid theory.  One makes a phase-space modification to the Hamiltonian~(\ref{H-FL}), keeps the
evolution equation~(\ref{H-evol}) unchanged, but alters the commutator
of $n_\p(\x)$ to be consistent with Eqs.~(\ref{PB}).  We shall now
work out this commutator.

Let us assume that there are two operators $\hat A$ and $\hat B$ defined as
\begin{equation}
  \hat A = \int\!{\rm d}\xi\, \sqrt{\omega}\, A(\xi)n(\xi),\quad
  \hat B = \int\!{\rm d}\xi\, \sqrt{\omega}\, B(\xi)n(\xi).
\end{equation}
Then it seems natural to define the commutator between $n(\xi)$ so that
\begin{equation}\label{comm-naive}
  [\hat A,\, \hat B] = -{\rm i}\!\int\!{\rm d}\xi\, \sqrt{\omega}\, \omega^{ab}
  \d_a A \d_b B\, n(\xi).
\end{equation}
This form, however, is deficient in one respect: it makes use of
$n(\xi)$ in the whole Fermi volume, while we expect the physics to be
concentrated near the Fermi surface.  We shall therefore postulate
another form for the commutator,
\begin{equation}\label{comm-Berry}
  [\hat A,\, \hat B] = 
  -\frac{\rm i}2\!\int\!{\rm d}\xi\, \sqrt{\omega}\, \omega^{ab}
  (A\d_a B- B \d_a A ) \d_b n(\xi).
\end{equation}
If we integrate by part in this equation, using
$\d_b(\sqrt{\omega}\,\omega^{ab})=0$ (a consequence of the Bianchi
identity), we bring Eq.~(\ref{comm-Berry}) into the form of
(\ref{comm-naive}).  However, now the commutator depends only on the
physics near the Fermi surface.  Moreover, we may have problems
defining the integral in Eq.~(\ref{comm-naive}) when the Berry
curvature is singular inside the Fermi volume (as in the case when the
Berry curvature flux is nonzero), while Eq.~(\ref{comm-Berry}) is
completely well-defined in this case.  We will take
Eq.~(\ref{comm-Berry}) as the equation defining the commutators of the
occupation number operator.

It is possible to write down explicitly the commutator $[n_\p(\x),\,
  n_\q(\y)]$. We shall not do it here.  Instead, we shall notice that
if $A$ and $B$ are not linear in $n(\xi)$, but are instead general
functionals of $n(\xi)$, then the commutator between them can still be
computed explicitly,
\begin{multline}\label{comm-Berry-2}
  [\hat A,\, \hat B] = 
  -\frac{\rm i}2\!\int\!{\rm d}\xi\, \sqrt{\omega}\, \omega^{ab}
  \biggl(\frac{\delta\hat A}{\delta n(\xi)}\d_a \frac{\delta\hat B}{\delta n(\xi)}
  - \\
  \frac{\delta\hat B}{\delta n(\xi)}\d_a \frac{\delta\hat A}{\delta n(\xi)}
  \biggr) 
  \d_b n(\xi).
\end{multline}
Equation~(\ref{comm-Berry-2}) is particularly useful when $\hat A$ is the
Hamiltonian, for which we know that $\delta H/\delta
n_\p(\x)=\epsilon_\p(\x)$.

\emph{Commutator of density operator.}---%
We now show that the Berry curvature leads to an anomalous term in the
equal-time commutator of the density operator $n(\x)$ at two points.
Moreover, if the Berry curvature has a nonzero magnetic flux through
the Fermi sphere, then the commutator has a contribution from the 
external magnetic field,
\begin{equation}\label{nn-comm}
  [ n(\x),\, n(\y)] = -\mathrm{i}\left(\bm{\nabla}\times\bm{\sigma}+
    \frac k{4\pi^2} {\bf B}\right)\cdot\bm{\nabla}\delta(\x-\y),
\end{equation}
where $\bm{\sigma}$ is defined as
\begin{equation}\label{sigma}
  \sigma_i(\x) = -\!\int\!\frac{{\rm d}\p}{(2\pi)^3}\,
    p_i\Omega_k \frac{\d n_\p(\x)}{\d p_k}\,,
\end{equation}
and $k$ is the monopole charge inside the Fermi surface,
\begin{equation}
  k = \frac1{2\pi}\!\int\!{\rm d}{\bf S} \cdot \bm{\Omega}.
\end{equation}
We note that both $\bm{\sigma}$ and $k$ involve
only the physics near the Fermi surface.

To derive Eq.~(\ref{nn-comm}), first we write
the density operator as
\begin{equation}
  n(\y) =\! \int\!\frac{{\rm d}\p}{(2\pi)^3}\,(1+{\bf B}\cdot\bm{\Omega}) 
  n_\p (\y) 
  = \!\int\!{\rm d}\Gamma\, \delta(\x-\y) n_\p(\x) .
\end{equation}
The commutator of the density operator at two different points is, according
to Eq.~(\ref{comm-Berry}),
\begin{multline}
  [ n(\y),\, n(\z) ] = -\frac{\rm i}2\!\int\!
  {\rm d}\Gamma\,  \delta( \x-\y)\d_i\delta(\x-\z) \biggl[ \{x_i,\, x_j\} 
   \frac{\d n_\p}{\d x_j} \\
  +  \{x_i,\, p_j\} \frac{\d n_\p}{\d p_j} 
   \biggr] - (\y\leftrightarrow\z).
\end{multline}
The $\{x_i,\, x_j\}$ term in the commutator is reduced to
\begin{equation}
  - {\rm i} \d_i\delta(\y-\z)\!\int\!\frac{{\rm d}\p}{(2\pi)^3}\,
    \epsilon_{ijk} \Omega_k \frac{\d n_\p}{\d x_j}
  = - {\rm i} (\bm{\nabla}{\times}\bm{\sigma}) \cdot \bm{\nabla}
      \delta(\y-\z),
\end{equation}
where $\bm{\sigma}$ is defined in Eq.~(\ref{sigma}). 
The $\{x_i,\, p_j\}$ term in the commutator can be rewritten as
\begin{equation}\label{comm-integral}
  {\rm i}B_i\d_i\delta(\y-\z)\!\int\!\frac{{\rm d}\p}{(2\pi)^3}\,
  \Omega_j \frac{\d n_\p}{\d p_j} \,.
\end{equation}
and, by integration by part, taking into account $\d_i \Omega_i=0$ around the Fermi surface, $n_\p=1$ deep inside the Fermi surface and $n_\p=0$ far
outside the Fermi sphere, it becomes
\begin{equation}
  -{\rm i}\frac k{4\pi^2} {\bf B}\cdot \bm{\nabla}\delta(\y-\z).
\end{equation}
Combining two contributions, we find Eq.~(\ref{nn-comm}).

\emph{From density-density commutator to anomalous 
nonconservation of current.}---%
The connection between the anomalous density-density commutator [the
term proportional to ${\bf B}$ in Eq.~(\ref{nn-comm})] and triangle
anomalies is known in the context of relativistic quantum field
theory~\cite{Jackiw,Faddeev:1984jp}.  Here we derive this connection
using the Hamiltonian formalism and show how the anomalous Hall
current and the triangle anomaly can be traced to the two
contributions to the density-density commutator.

Let us first assume that our system is in a static magnetic field, but
the electric field is turned off.  In this case, the system is
described by the Hamiltonian~(\ref{H-FL}), and by commuting the
Hamiltonian with the particle number operator $n(\x)$, the continuity equation can be derived, 
\begin{equation}
  \dot n = {\rm i}[H,\, n] = -\bm{\nabla}\cdot{\bf j},
\end{equation}
where the particle number current ${\bf j}$ is
\begin{equation}\label{ji}
  {\bf j} = \!\int\!\frac{{\rm d}\p}{(2\pi)^3} \biggl[ 
    - \epsilon_\p \frac{\d n_\p}{\d\p}
    - \Bigl(\bm{\Omega}\cdot \frac{\d n_\p}{\d\p}\Bigr)\epsilon_\p {\bf B}
    - \epsilon_\p\, \bm{\Omega}\times \frac{\d n_\p}{\d \x}\biggr].
\end{equation}
Note that by integration by part, the first term in the brackets in the
right-hand side of Eq.~(\ref{ji}) can be written in the familiar form
$n_\p {\bf v}$, where ${\bf v}=\d\epsilon_\p/\d\p$.  This would be the
only term in the current in the absence of Berry curvature.

Now we turn on a static electric field by putting the system in an external scalar potential $\phi(\x)$, ${\bf E}=-\bm{\nabla}\phi$.  The Hamiltonian is now
\begin{equation}
  H' = H +\! \int\!{\rm d}\x\, \phi(\x) n(\x).
\end{equation}
The added term does not commute with $n$ and changes the time
evolution of the latter,
\begin{equation}
  \d_t n(\x) = {\rm i}[H',\, n(\x)] =
  -\bm{\nabla}\cdot {\bf j} - \left(
   \bm{\nabla}{\times} \bm{\sigma}+ \frac k{4\pi^2} {\bf B}\right)
   \cdot\bm{\nabla}\phi(\x).
\end{equation}
This equation can be rewritten as
\begin{equation}\label{eq-anom}
  \d_t n + \bm{\nabla}\cdot{\bf j}' = \frac k{4\pi^2} {\bf E}\cdot {\bf B},
\end{equation}
where
\begin{equation}\label{AHE}
  {\bf j}' = {\bf j} + {\bf E}\times \bm{\sigma}.
\end{equation}
The second term in Eq.~(\ref{AHE}) is the usual anomalous Hall
current.  On the other hand, Eq.~(\ref{eq-anom}) implies that the
particle number around the Fermi surface is not conserved when both
electric and magnetic fields are turned on.  This is the effect of
triangle anomalies in quantum field theory. For example, relativistic
right-handed free fermions have $k=1$, and left-handed free fermions
have $k=-1$.  Here we show that this effect depends only on the
monopole charge of the Berry curvature on the Fermi surface and is
not modified by interactions.  Since the total charge is conserved, 
all different contributions to the current
nonconservation should sum up to zero.

\emph{Chiral magnetic effect}---%
Let us compute the current, given by Eq.~(\ref{ji}), in the thermal
equilibrium state,
where quasiparticles have a Fermi-Dirac distribution function,
\begin{equation}
  n_\p = f(x) = \frac1{e^x+1}\,, \quad x=\frac{\epsilon_\p-\mu}T\,.
\end{equation}

There are three contributions to ${\bf j}$ corresponding to three terms
in the brackets in the right-hand side of Eq.~(\ref{ji}).  The third
term involves spatial derivatives and vanishes in the ground state.
We now show that the first term also vanishes identically.  For this
end it is useful to introduce the function $g(x)=\!\int_{-\infty}^x\!{\rm d}y\, yf'(y)$,
for which $g(-\infty)=g(+\infty)=0$.  Then
\begin{equation}
  -\!\int\!\frac{{\rm d}\p}{(2\pi)^3}\, \epsilon_\p \frac{\d n_\p}{\d \p}
  = - \!\int\!\frac{{\rm d}\p}{(2\pi)^3}\biggl[\mu\frac{\d n_\p}{\d \p}
    + T\frac{\d}{\d \p} g(x)\biggr] = 0.
\end{equation}
Similarly, the second contribution can be written as
\begin{equation}
  -{\bf B}\!\int\!\frac{{\rm d}\p}{(2\pi)^3}\bm{\Omega}\cdot\biggl[
    \mu\frac{\d n_\p}{\d\p} +
   T \frac\d{\d\p} g(x)
   \biggr],
\end{equation}
and the integrals can be evaluated as in
Eq.~(\ref{comm-integral}).  As the result, we find
\begin{equation}
  {\bf j} = \frac k{4\pi^2}\mu{\bf B}.
\end{equation}
Let us assume for definiteness that there are two Fermi surfaces with
$k=1$ and $k=-1$.  If the two Fermi surfaces have equal chemical
potential, then the total current is equal to 0.  However, if the
chemical potentials are unequal (which can be achieved by turning on
an ${\bf E}\cdot{\bf B}$ for a finite time), then there will be a
current equal to $(\mu_+-\mu_-){\bf B}/4\pi^2$ in this
quasiequilibrium state.  This is the chiral magnetic
effect~\cite{Nielsen:1983rb,Fukushima:2008xe}.

\emph{Conclusion.}---%
The calculation above ties the anomalous current nonconservation to a
property of the Fermi surface only (the Berry curvature) and hence can
be applied to Fermi liquids, even when the interaction between
original fermions is strong.  This is done by using a kinetic equation
for quasiparticles; a more microscopic derivation of this
equation~\cite{ShindouBalents,WongTserkovnyak} is desirable.

It would be interesting to explore further physical consequences of
Berry curvature on Landau Fermi liquid theory. Particularly
interesting are the effects of Berry curvature on the collective modes
and on the response of the Fermi liquids.  It is also interesting to
include the collision term into the kinetic equation and investigate
the hydrodynamic regime.  In relativistically invariant theories, the
effects of triangle anomalies have been investigated, both within
hydrodynamics and by using gauge-gravity
duality~\cite{Son:2009tf,Neiman:2010zi,Landsteiner:2011cp}, and there
have been attempts to derive hydrodynamics from kinetic
theory~\cite{Loganayagam:2012pz}.  The kinetic approach allows us to
go beyond the hydrodynamic regime and beyond systems with relativistic
invariance.  On the other hand, some interesting phenomena associated
with anomalies in relativistic theories, like the Alfv\'en-type modes
propagating along the direction of the magnetic
field~\cite{Newman:2005hd,Kharzeev:2010gd}, may be directly
investigated using the kinetic equation in the condensed-matter
context.

Finally, the understanding obtained here should allow one to formulate
the criteria of anomalies matching for dense states matter, i.e.,
quark matter phases with Fermi surfaces.

The authors thank L.~Balents, R.~Loganayagam, B.~Spivak, P.~Sur\'owka,
and A.~Vishwanath for discussions. This work is supported, in part, by
DOE Grant No.\ DE-FG02-00ER41132. N.Y. is supported, in part, by JSPS
Postdoctoral Fellowships for Research Abroad.

\end{document}